\documentclass{mn2e}
\usepackage{times}
\usepackage{epsf}
\newcommand{\etal}{{et al}\/.}
\begin{document}
\title[High-frequency polarimetric observations of 3C\,171]{Probing the extended
emission-line region in 3C\,171 with high-frequency radio polarimetry}
\author[M.J.~Hardcastle]{M.J.\ Hardcastle\footnotemark\\
Department of Physics, University of Bristol, Tyndall Avenue,
Bristol BS8 1TL}
\maketitle
\begin{abstract}
I present high-frequency polarimetric radio observations of the radio
galaxy 3C\,171, in which depolarization is associated with an extended
emission-line region. The radio hotspots, known to be depolarized at
low radio frequencies, become significantly polarized above
(observer's-frame) frequencies of 15 GHz. There is some evidence that
some of the Faraday rotation structure associated with the
emission-line regions is being resolved at the highest resolutions
(0.1 arcsec, or 400 pc); however, the majority remains unresolved.
Using the new radio data and a simple model for the nature of the
depolarizing screen, it is possible to place some constraints on the
nature of the medium responsible for the depolarization. I argue that
it is most likely that the depolarization is due not to the
emission-line material itself but to a second, less dense, hot phase of the
shocked ISM, and derive some limits on its density and temperature.
\end{abstract}
\begin{keywords}
galaxies: active -- galaxies: individual: 3C\,171 -- galaxies: ISM
\end{keywords}
\footnotetext{E-mail {\it m.hardcastle}@{\it bristol.ac.uk}}
\section{Introduction}

Extended emission-line regions (EELR) around radio galaxies are
common, and often aligned with the axes of the extended radio emission
(e.g.\ McCarthy \etal\ 1987; McCarthy 1993; McCarthy, Baum \& Spinrad
1996). There is considerable debate as to whether EELR are ionized by
photons from the nucleus or by shocks driven by the jets (e.g.\
Tadhunter \etal\ 1998). Most EELR around low-redshift sources can
adequately be described by central-illumination models, while sources
which show evidence for shock ionization (such as extreme
emission-line kinematics) are largely at high
redshift. However, there is evidence that a small minority of
low-redshift radio galaxies are also more adequately described by a
shock-ionization model. If the ionization is due to shocks, the
physical conditions in and kinematics of the EELR may give us
important information about the energetics of the radio source as a
whole.

One good candidate low-redshift shock-ionization source is 3C\,171, an
unusual $z=0.2384$ radio galaxy. Its large-scale radio structure has
been well studied (Heckman, van Breugel \& Miley 1984; Blundell 1996).
In its inner regions it is similar to a normal FRII, but
low-surface-brightness plumes instead of normal lobes extend north and
south from the hotspots. High-resolution images show knotty jets
connecting the radio core with the hotspots (Hardcastle \etal\ 1997).
3C\,171 is associated with a prominent EELR, elongated along the jet
axis. Heckman \etal\ found that the emission-line gas was brightest
near the hotspots, suggesting that the radio-emitting plasma is
responsible for powering the emission-line regions. More recently
Clark \etal\ (1998) have made WHT observations of the emission-line
regions and shown that their ionization states are most consistent
with a shock-ionization model. The unusually strong EELR in this
source may be indicative of an environment unusually rich in dense,
warm gas; the interactions between the jets and this gas might then be
responsible for the peculiar large-scale radio structure.

Heckman \etal\ pointed out the depolarization of the radio source near
the hotspots at frequencies up to 15 GHz. The strong spatial
relationship between the EELR (or the material traced by the EELR) and
depolarization was pointed out by Hardcastle \etal\ (1997), who
superposed an 8-GHz Very Large Array (VLA) image on the {\it Hubble
Space Telescope} ({\it HST}) snapshot (de Koff \etal\ 1996) in the
F702W filter (which at this redshift passes several strong lines). The
source has since been observed with a number of other {\it HST}
filters, and Fig.\ 1 shows a superposition of the 8-GHz map on an {\it
HST} image in the [OIII]495.9/500.7 nm lines. Regions of low
polarization in the radio map can be seen to correspond to the positions of
emission-line material. The eastern hot spot is almost completely
unpolarized at 8 GHz and below, while the western hot spot is
partially depolarized.

The close assocation between the emission-line regions and the radio
depolarization in 3C\,171 opens up the possibility of using radio data to
constrain unknown physical conditions, such as the electron number
density and the magnetic field strength in the gas surrounding the source.
In this paper, I present new radio observations of 3C\,171, and
discuss the constraints that they place on the nature of its environment.

Throughout the paper, I use a cosmology with $H_0 = 65$ km s$^{-1}$
Mpc$^{-1}$, $\Omega_{\rm m} = 0.3$ and $\Omega_\Lambda = 0.7$. At the
redshift of 3C\,171, 1 arcsec corresponds to 4.07 kpc.

\begin{figure*}
\epsfxsize 12cm
\epsfbox{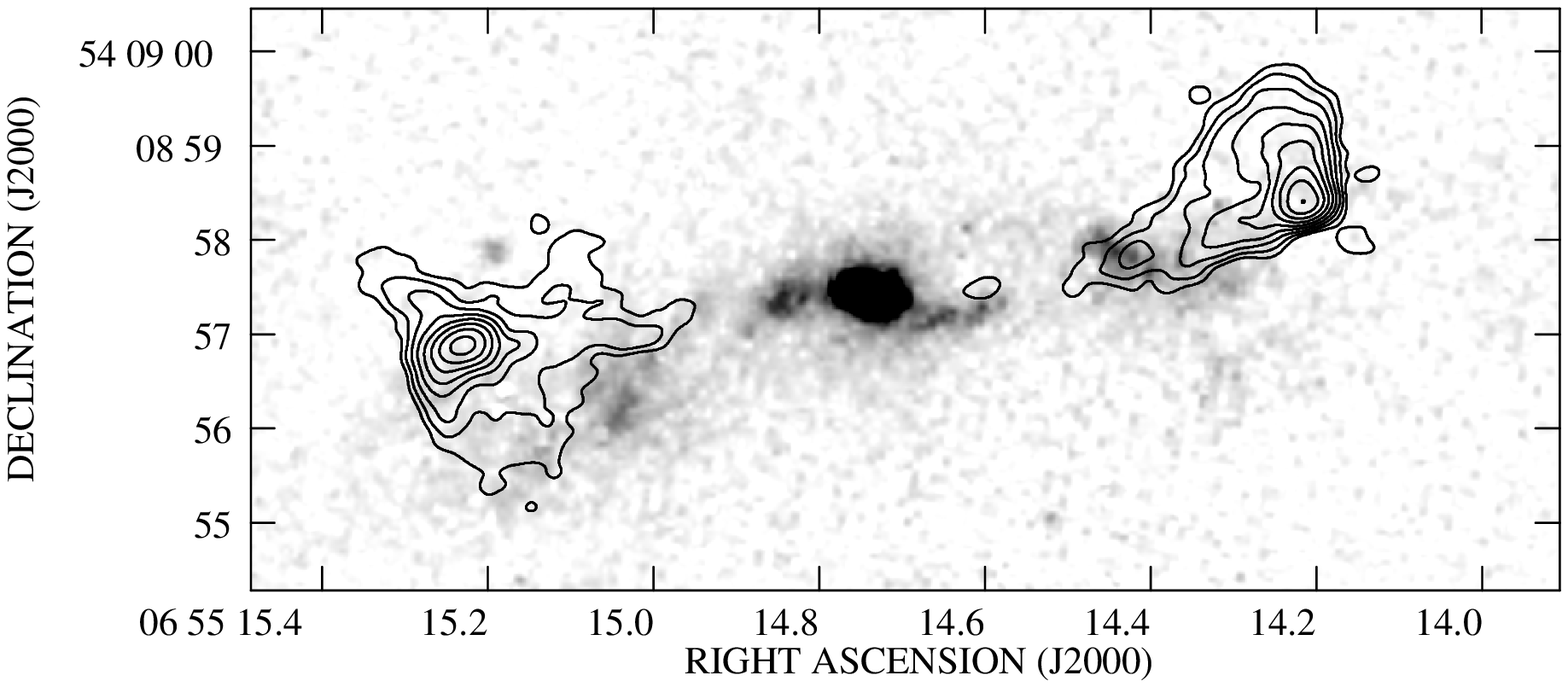}
\epsfxsize 12cm
\epsfbox{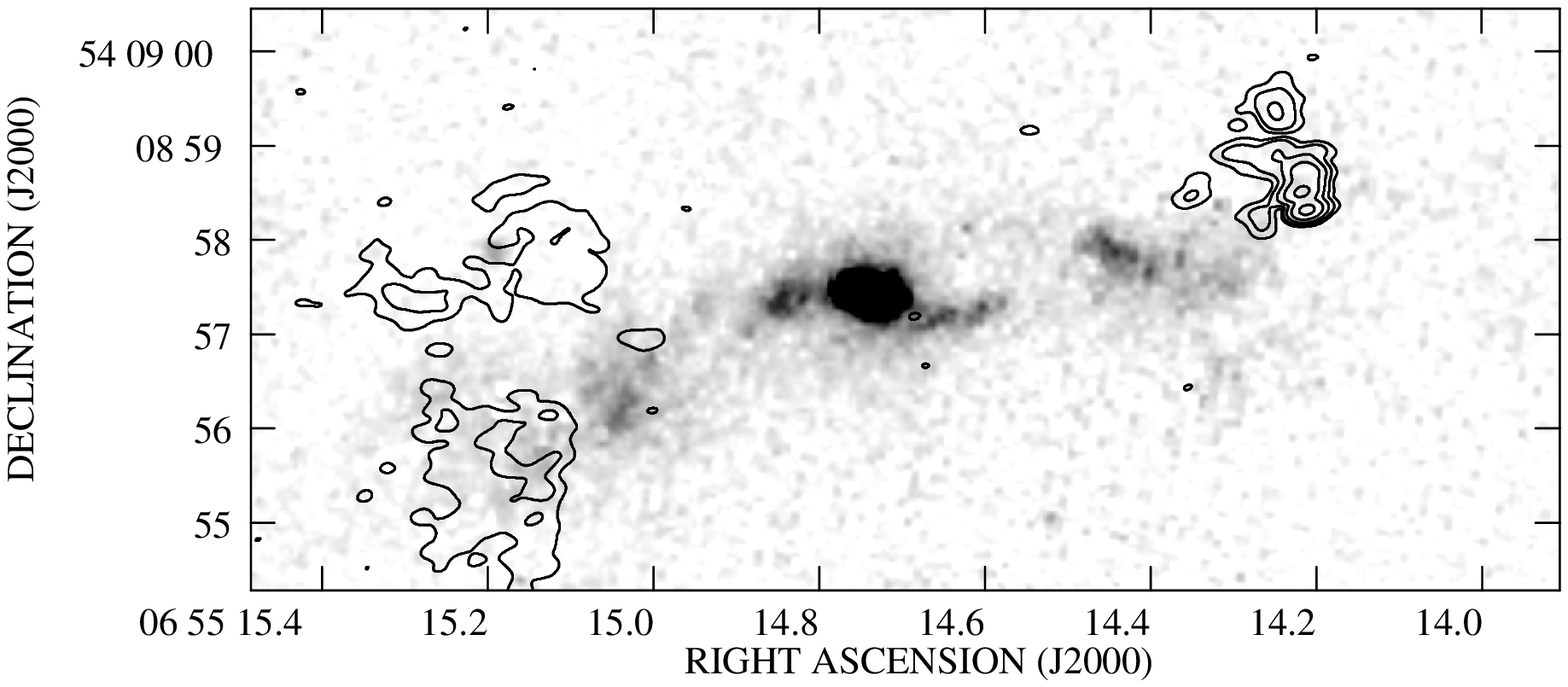}
\caption{8.1 GHz VLA image, made with VLA A-configuration data (see
the text). Above is the total intensity image (contours at $300 \times
(1, 2, 4\dots)$ $\mu$Jy beam$^{-1}$; below is the polarized intensity
image (contours at $150 \times (1, 2, 4\dots)$ $\mu$Jy
beam$^{-1}$. The restoring beam is an $0.29 \times 0.20$-arcsec
elliptical Gaussian in PA $-79$\degr. The greyscale shows a
3000-s exposure in the FR680 ramp filter of the {\it HST}'s
WFPC2, kindly supplied by Clive Tadhunter, and is dominated by emission
from the [OIII] 495.9/500.7 nm lines. Note the strong relationship between
depolarization and emission-line regions.}
\end{figure*}

\section{Observations and analysis}

The degree of depolarization is in general expected to be reduced at
higher frequencies and higher resolutions. As parts of 3C\,171's
hotspots had been found to be completely depolarized at the full
resolution of the VLA ($\sim 0.25$ arcsec) at 8.1 GHz, I elected to
observe the source with the A and B arrays of the VLA at the U and K
bands, $\sim 15$ and $\sim 22$ GHz. In addition, I made use of the
A-array 8.1-GHz observations discussed by Hardcastle \etal\ (1997).

Details of the observations used are listed in Table \ref{obstab}. The
new observations were carried out in fast-switching mode\footnote{See
http://info.aoc.nrao.edu/vla/html/highfreq/hffastswitch.html.} with an
integration time of 3.3 s, using frequent observations of a nearby
bright point-source calibrator, 0749+540. The cycle time used was 320
s, with 80 s on calibrator and 240s on source, giving typical times of
50 s on the calibrator and 210 s on source after dead time due to
slewing. (The low efficiency was due to the lack of a good nearby
calibrator.) Primary referenced pointing\footnote{See
http://info.aoc.nrao.edu/vla/html/refpt.shtml.} was carried out, using
short X-band observations of 0749+540 taken every hour, to avoid loss
of sensitivity due to VLA pointing errors. The observations were made
with two nearby observing frequencies (Table \ref{obstab}) with a
50-MHz bandwidth in each. 3C\,286 was used as the flux and
polarization position angle calibrator. The data were reduced in a
standard manner within {\sc aips}. {\sc elint} was used to correct for
the dependence of antenna gain on elevation. The use of fast-switching
mode led to observations which were very well phase- and
amplitude-calibrated after the standard calibration procedure:
however, there was enough signal for a single iteration of phase
self-calibration for each dataset, which gave a marginal improvement
in image quality. Finally, the A- and B-array datasets were combined
using {\sc dbcon}, {\sc uvfix} was used to align the datasets at the
three wavelengths at the radio core, using the K-band dataset as the
reference, and final images were made using {\sc imagr}, combining the
data from the two observing frequencies in each band. This resulted in
images at effective frequencies of 8.09, 14.94 and 22.46 GHz. For
simplicity, these frequencies will be referred to as 8, 15 and 22 GHz
from now on.

\begin{table}
\caption{VLA observations}
\begin{tabular}{llrrlr}
\hline
Array&Band&\multicolumn{2}{c}{Observing frequencies}&Date&Integration\\
&&\multicolumn{2}{c}{(GHz)}&&time (h)\\
\hline
A&X&8.065&8.115&1995 Aug 06&1\\
A&U&14.915&14.965&2001 Jan 16&4\\
A&K&22.435&22.485&2001 Jan 16&4\\
B&U&14.915&14.965&2001 Apr 01&1\\
B&K&22.435&22.485&2001 Apr 01&3\\
\hline
\end{tabular}
\label{obstab}
\end{table}

\section{Results}

Initially, images at the full resolution of the data were made: the
image properties are listed in Table \ref{maps}. These images were
made using a matched shortest baseline of 18 k$\lambda$, and so are in
principle sensitive to structures on the same angular scales: in
practice, however, large-scale structure is resolved out at higher
frequencies, and the steep spectrum of the source at high frequencies
(Blundell 1996) also reduces the probability of detecting the lobes at
15 and 22 GHz.

\begin{table}
\caption{Resolutions and off-source noise levels of radio images}
\label{maps}
\begin{tabular}{llrrr}
\hline
Frequency&Image&\multicolumn{3}{c}{Noise ($\mu$Jy beam$^{-1}$)}\\
(GHz)&&Stokes $I$&Stokes $Q$&Stokes $U$\\
\hline
8&$0.29 \times 0.20$, $-79$\degr&54&33&33\\[4pt]
15&$0.14 \times 0.12$, $-87$\degr&49&47&46\\
&$0.30 \times 0.20$, $-80$\degr&76&44&42\\
&(matched to 8-GHz)\\[4pt]
22&$0.10 \times 0.09$, $-81$\degr&37&33&33\\
&$0.14 \times 0.12$, $-87$\degr&40&28&27\\
&(matched to 15-GHz)\\
&$0.30 \times 0.20$, $-80$\degr&74&44&42\\
&(matched to 8-GHz)\\
\hline
\end{tabular}
\end{table}

\begin{figure*}
\hbox{
\epsfxsize 7cm
\epsfbox{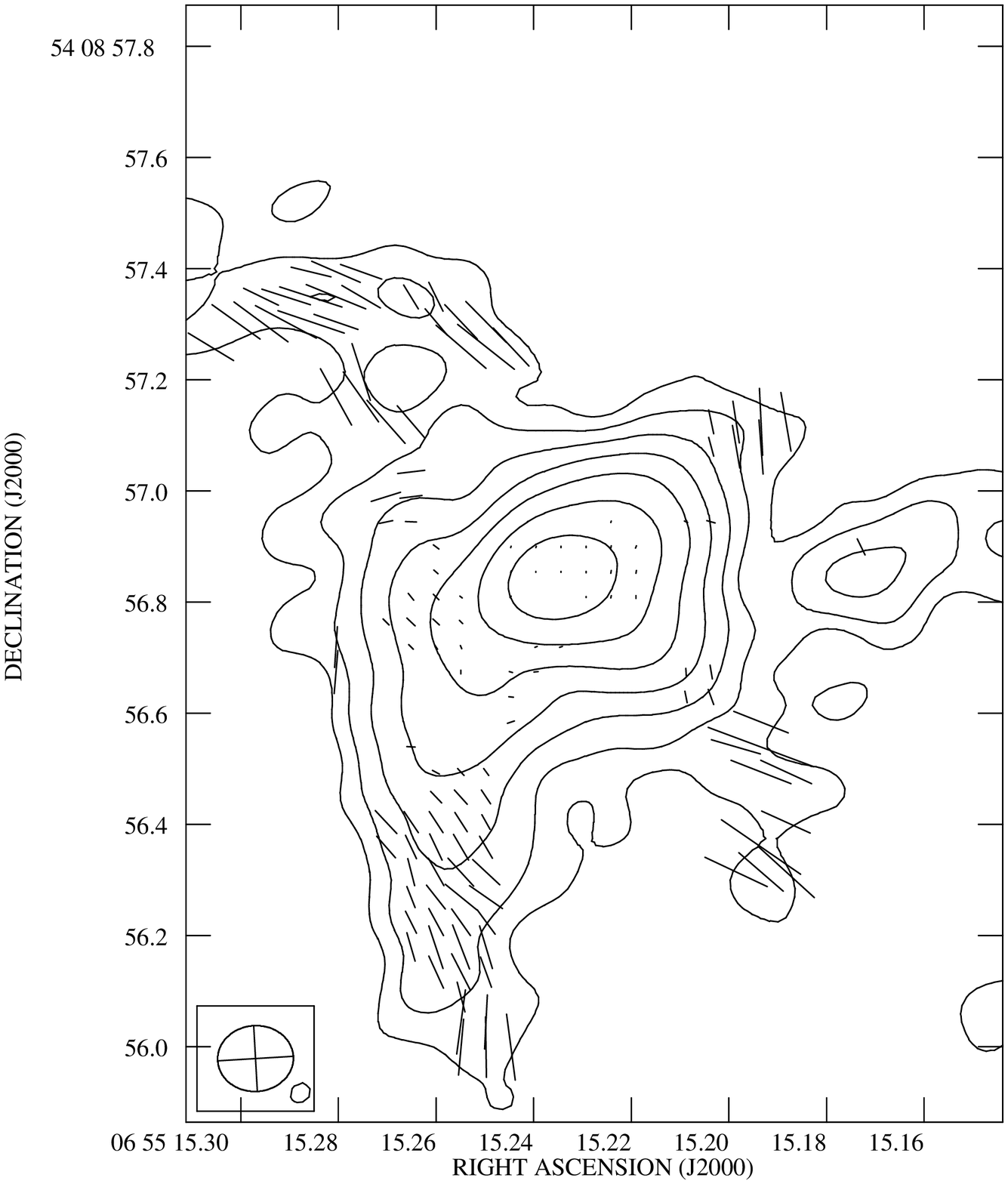}
\epsfxsize 7cm
\epsfbox{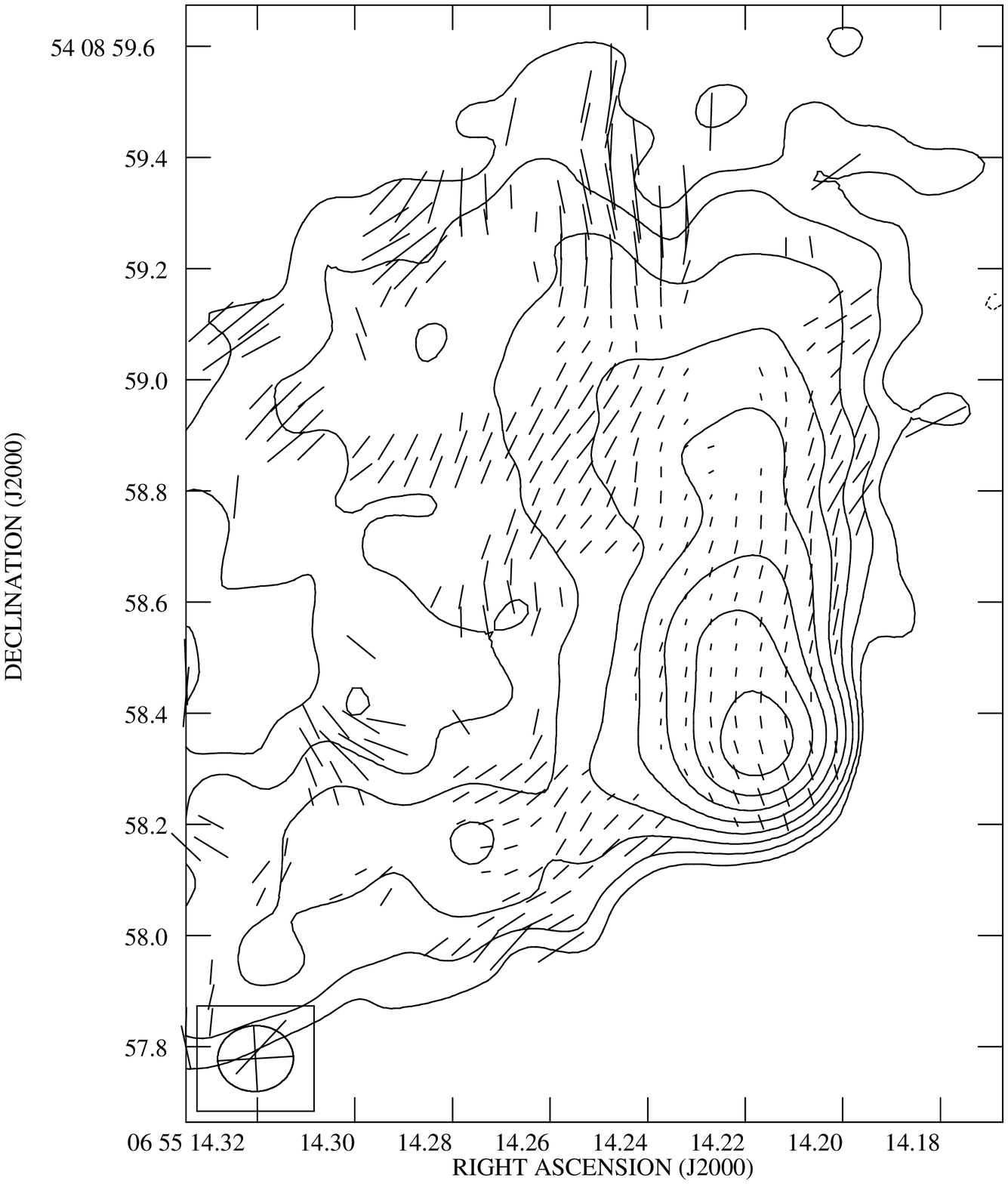}
}
\hbox{
\epsfxsize 7cm
\epsfbox{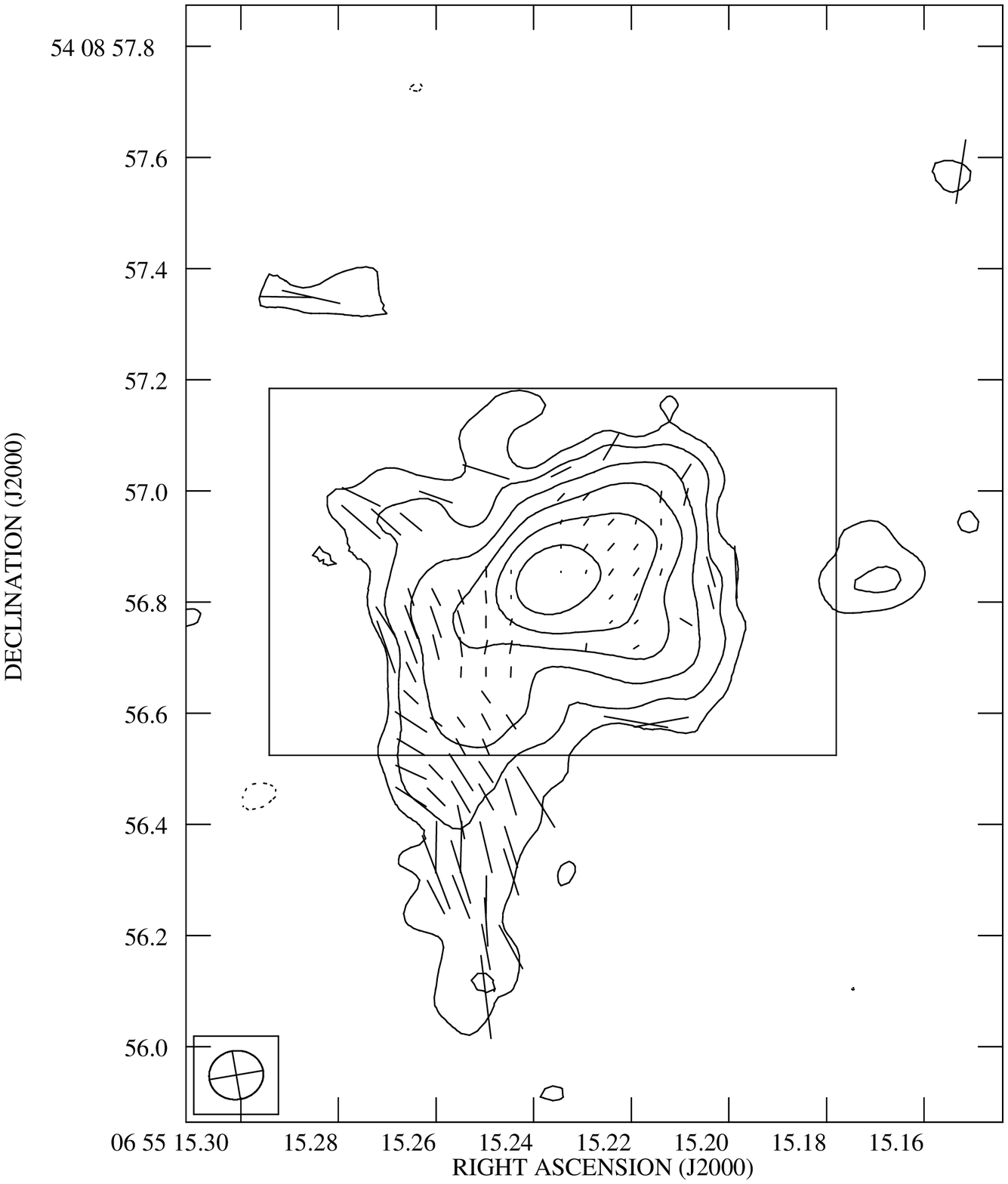}
\epsfxsize 7cm
\epsfbox{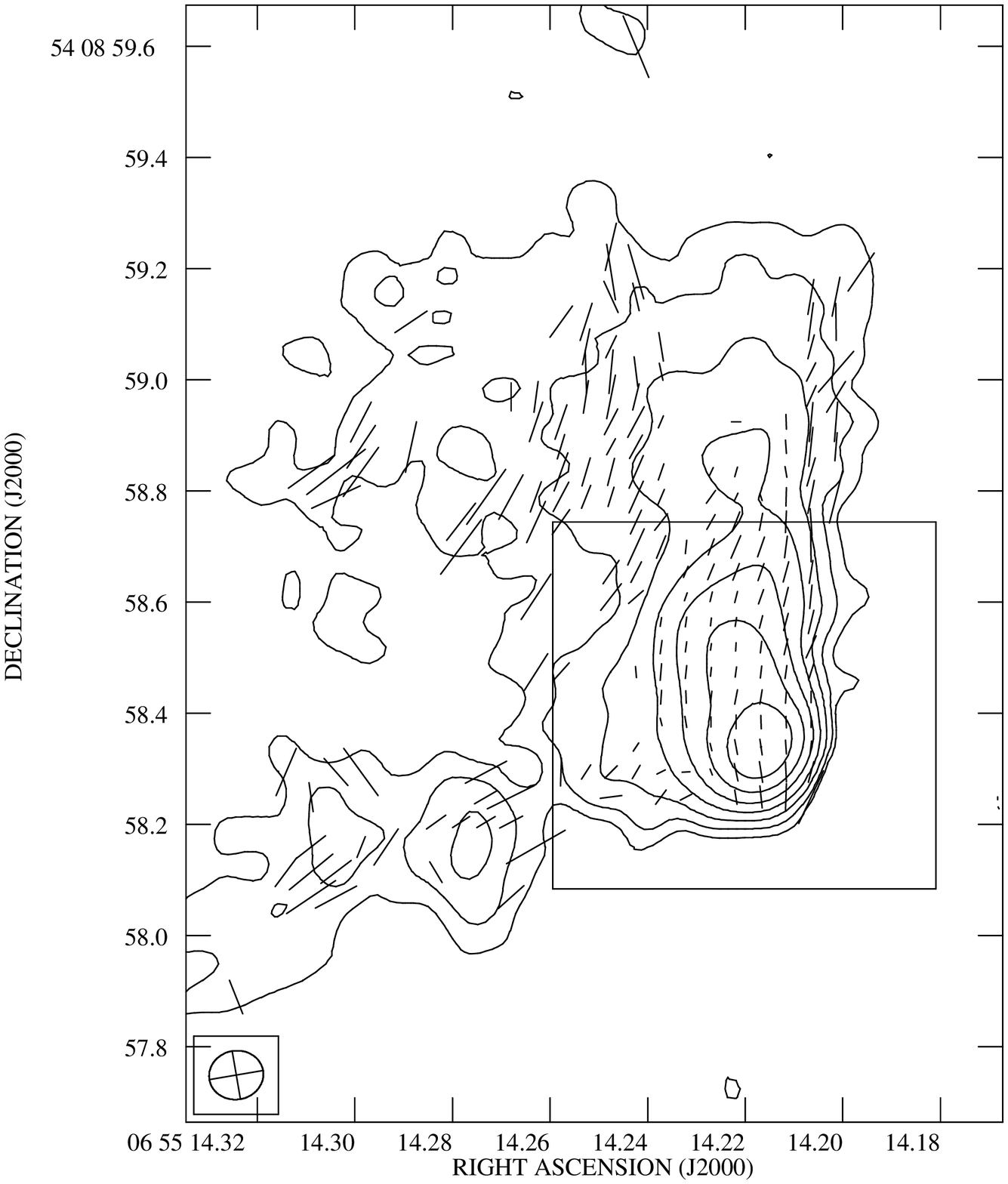}
}
\caption{The hotspots of 3C\,171 at full resolution  at 15 GHz (top,
$0.14 \times 0.12$ arcsec)
and 22 GHz (bottom, $0.10 \times
0.09$ arcsec). The off-source noise levels in the two images are given in Table
\ref{maps}. The lowest contour level is the $3\sigma$ level and the
contours are logarithmic, increasing by a factor 2. Negative contours
are dashed. Polarization vectors are perpendicular to the direction of
the $E$-vector, and so would show magnetic field direction in the
absence of Faraday rotation. Their lengths indicate relative
fractional polarization. Polarization vectors are oversampled by a
factor $\sim 2$ for visibility. Boxes on the lower panels show the
regions used for integration of total and polarized intensity: see the
text.}
\label{22hotsp}
\end{figure*}

In total intensity, the differences between the images are therefore
predictable. The radio core is detected at all three frequencies, with
flux densities of 1.8, 3.4 and 2.5 mJy at 8, 15 and 22 GHz
respectively, giving it a steep spectrum at the higher frequencies.
The only particularly notable feature to emerge from the new data is
the compactness of the hotspots, which are only just resolved at the
full resolution at 22 GHz, and are therefore only tens of milliarcsec
(hundreds of pc) in size (Fig. \ref{22hotsp}). The W hotspot shows a
well-resolved extension to the N, which may be tracing the current
direction of outflow from the hotspot, while the E hotspot shows a
striking bow-shock-shaped extension around the central compact
component, superposed on a N-S filament. The hotspots have steep
spectra at these high frequencies: the integrated spectral indices in
the hotspots are $\alpha_{8}^{15} = 0.92$ and $\alpha_{15}^{22} =
1.38$ (W hotspot) and $\alpha_{8}^{15} = 1.08$ and $\alpha_{15}^{22} =
1.58$ (E hotspot).

In polarization, however, the data clearly show the E hotspot to be
polarized at 15 and 22 GHz, although it is essentially unpolarized at
8 GHz and below. The partial depolarization of the W hotspot also
disappears at high frequencies. Some of this might be thought to be
due to simple beam depolarization in the low-resolution 8-GHz images:
there clearly is some polarization structure in the E hotspot in the
images of Fig.\ \ref{22hotsp}. To investigate this, I made images at
lower resolutions, matched to the resolution of the 8-GHz data, using
appropriate weighting and tapering of the $u, v$ plane, and compared
the integrated total and polarized fluxes of the two hotspots at
different resolutions using the AIPS verb {\sc imstat}, measuring
using boxes which are shown in Fig.\ \ref{22hotsp}. Beam
depolarization effects should be independent of frequency, to a first
approximation, although the effects of spectral index mean that we are
not seeing exactly the same structures at each frequency: considering
the hotspot regions only should minimize these effects. Depolarization
due to a Faraday-rotating medium should have a strong frequency
dependence. The results of the measurements are given in Table
\ref{fltab}.

It can be seen from these measurements that the dominant effect is
frequency-dependent. The fractional polarization of the two hotspots
increases with increasing frequency even in images made with matched
resolutions. The total polarized flux remains approximately constant
with frequency in the W hotspot and increases with frequency in the E hotspot.
Since the flux measured in total intensity falls steeply with
frequency, we can be certain that this is not simply a spectral index
effect: a contaminating, highly-polarized component would have to have
a flat or inverted spectral index. There is some evidence for beam
depolarization effects in the differences between the fractional
polarizations observed at different resolutions with the same
frequency, particularly at 22 GHz, but this is clearly not the most
significant effect.

\begin{figure*}
\hbox{
\epsfxsize 7cm
\epsfbox{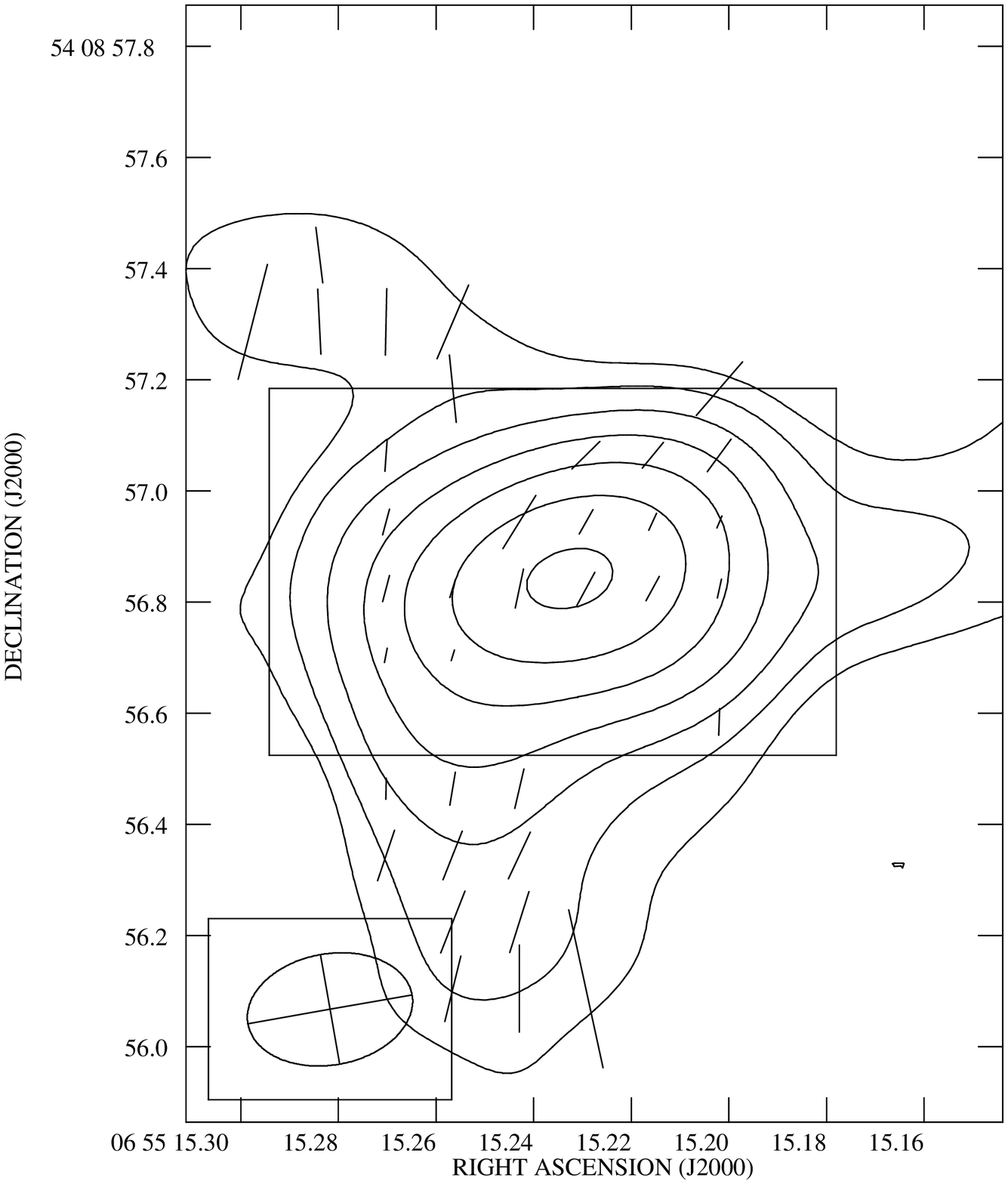}
\epsfxsize 7cm
\epsfbox{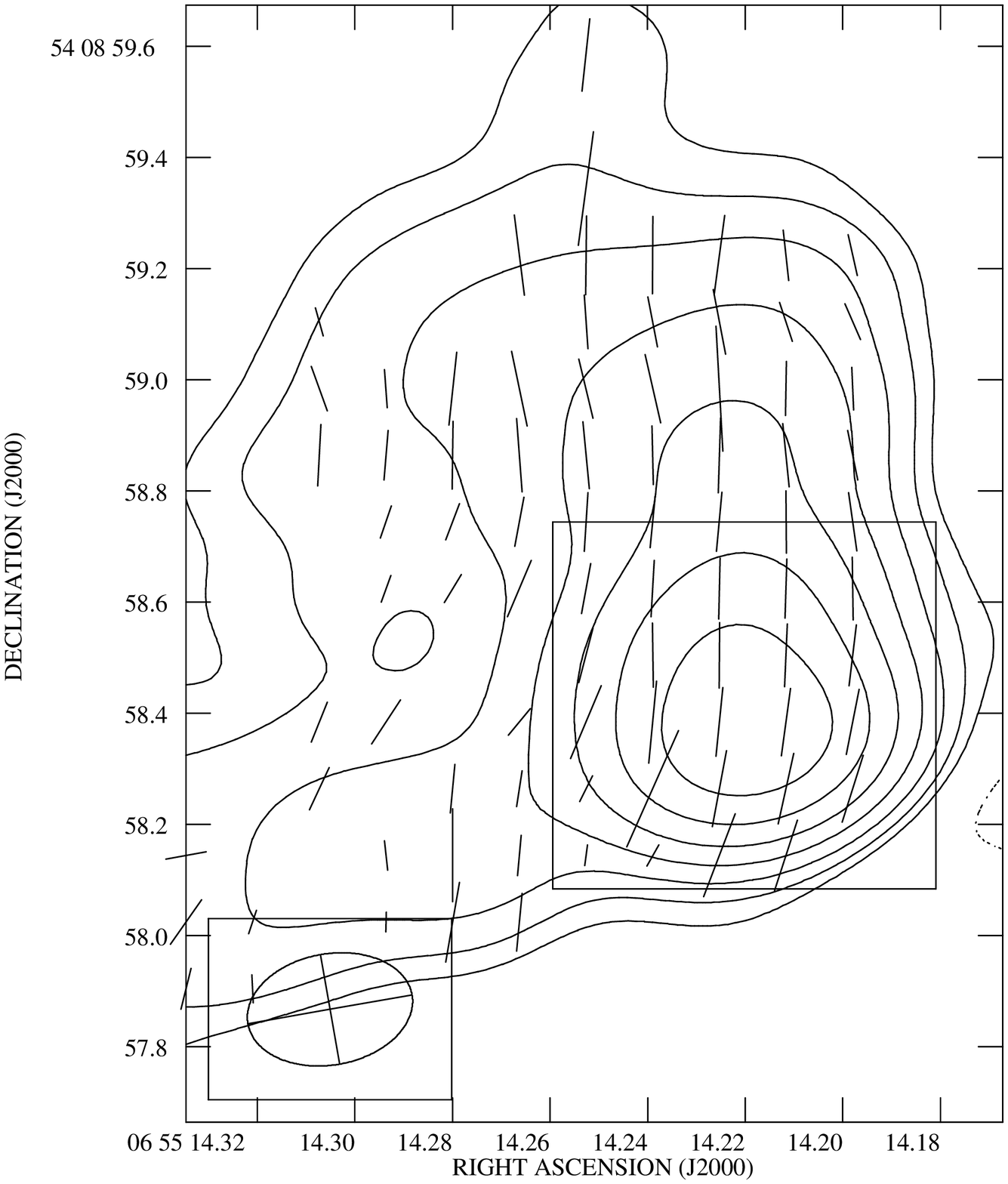}
}
\caption{Depolarization and rotation between 15 and 22 GHz in the
hotspots at $0.3 \times 0.2$ arcsec resolution. The vector lengths
show the ratio between the fractional polarizations at 15 and 22 GHz,
with shorter vectors implying more depolarization. The longest vectors
in both images correspond to a ratio of 1 (i.e., no depolarization).
The vector direction shows the rotation between the two frequencies,
with a vertical vector meaning no rotation, and positive rotations
being counterclockwise. Only points with $3\sigma$ detections of total
and polarized flux at both frequencies are plotted. Vectors are
oversampled by a factor $\sim 2$.}
\label{depol}
\end{figure*}

There is some limited evidence for Faraday rotation associated with
the region of depolarization. This is illustrated in Fig.\
\ref{depol}, which shows the depolarization and rotation between 15
and 22 GHz at a resolution matched to the 8-GHz map. The most
significant rotation is in the N part of the E hotspot. However, there
is no particularly strong relationship between Faraday rotation and
depolarization, and the rotation is never particularly large (modulo
possible 180\degr\ rotation effects).

\begin{table}
\caption{Total and polarized flux measurements of the E and W
hotspots}
\label{fltab}
\begin{tabular}{lllrrr}
\hline
Hotspot&Freq.&Image&Total flux&Polarized&Fractional\\
&(GHz)&&(mJy)&flux (mJy)&polarization\\
&&&&&(per cent)\\
\hline
W&8&Full&149.0&$6.5 \pm 0.1$&$4.4 \pm 0.1$\\[2pt]
&15&8&84.5&$9.6 \pm 0.1$&$11.4 \pm 0.1$\\
&15&Full&86.1&$10.3 \pm 0.2$&$12.0 \pm 0.2$\\[2pt]
&22&8&48.2&$6.7 \pm 0.1$&$13.9 \pm 0.2$\\
&22&15&49.4&$7.5 \pm 0.1$&$15.1 \pm 0.2$\\
&22&Full&49.2&$7.7 \pm 0.2$&$15.7 \pm 0.3$\\[4pt]
E&8&Full&116.5&$0.6 \pm 0.1$&$0.5 \pm 0.1$\\[2pt]
&15&8&60.1&$1.4 \pm 0.1$&$2.3 \pm 0.2$\\
&15&Full&60.9&$1.7 \pm 0.2$&$2.7 \pm 0.3$\\[2pt]
&22&8&31.6&$1.8 \pm 0.1$&$5.7 \pm 0.3$\\
&22&15&32.6&$2.8 \pm 0.1$&$8.2 \pm 0.4$\\
&22&Full&32.3&$2.7 \pm 0.2$&$8.6 \pm 0.6$\\
\hline
\end{tabular}
\begin{minipage}{\linewidth}
The `Image' column describes the resolution of the image used: `Full' denotes
the full resolution of the dataset, `8' means matched to the 8-GHz
dataset, and `15' means that the image was matched to the 15-GHz
dataset (the images are those described in Table \ref{maps}). All images
were made with matched shortest baselines. The errors on the polarized
intensity are determined from the off-source r.m.s. noise in the
Stokes $Q$ and $U$ images. Errors on the total intensity are similar in
magnitude, but are not quoted as they are not a significant
contributor to the errors in fractional polarization.
\end{minipage}
\end{table}

\section{Interpretation}

We can begin by making the assumption that the depolarizing medium is
external to the source. This is justified by the extremely good,
though not perfect, association between the EELR and the
depolarization. No model places the EELR inside the radio source -- it
is spatially quite distinct from, and physically larger than, the
radio emission at various points (Fig.\ 1) -- and at the same time it
is clearly related to the depolarization. We can therefore treat the
depolarizing material as an external screen of dimensions comparable
to those of the EELR.

The situation is then a classical problem in the theory of radio
polarization, discussed by Burn (1966). If a screen is external, two
extreme cases can be identified. In the first, the Faraday screen is
fully resolved by the telescope beam: in this case, no depolarization
should be observed, but Faraday rotation should be seen. Some sources
in rich cluster environments appear to be in this regime, in that the
Faraday screen is partially or fully resolved (e.g.\ Perley \& Carilli
1996). In 3C\,171, however, we see significant depolarization without
much associated Faraday rotation, as Fig.\ \ref{depol} shows. This is
likely to be due to a largely unresolved foreground screen. In the
limit that the size scale $d$ of the individual depolarizing regions
(regions of uniform magnetic field) is much smaller than the
resolution element, {\it and is well defined}, and that there are many
regions along each line of sight, the degree of polarization $p$ of a
part of the source seen through such a screen at a given frequency
$\nu$ is given by
\begin{equation}
p = p_i
\exp\left[-2K^2(n_e B_\parallel)_f^2 d R c^4\nu^{-4}\right]
\label{depeq}
\end{equation}
(Burn 1966) where $p_i$ is the intrinsic
polarization, $K$ is a constant with value $8.12 \times 10^{-3}$ rad
nT$^{-1}$ m kpc$^{-1}$, $(n_e B_\parallel)_f^2$ is the dispersion in the
product of number density and magnetic field strength and $R$ is the
line-of-sight depth through the medium.

We can try to apply this model to the observations, but there are a
number of reasons why it should be treated with caution. The first is
a practical one: the measurements of the polarized intensity of the
hotspots clearly include regions with different degrees of
polarization (Fig.\ \ref{depol}) and so, presumably, different
physical conditions in the depolarizing screen. At best we will obtain
some emission-weighted average of the physical conditions. To reduce
this effect, I consider below only the E hotspot, which exhibits more
uniform depolarization. The model is also not realistic. It assumes a
single physical size scale for the variations in Faraday depth,
parametrized by $d$. As Laing (1984) points out, realistic screens
will have a range of scale sizes, which tends to wash out the sharp
cutoff in polarization at low frequencies. This is probably the case
for our observations: the degree of polarization at 8 GHz, though low,
is still too high to be fitted with a simple model of the form of eq.\
\ref{depeq}. If we write eq.\ \ref{depeq} as $p = p_i
\exp(-C\nu^{-4})$, then we can solve for $C$ and $p_i$ using the 15
and 22 GHz data at $0.3 \times 0.2$-arcsec resolution, obtaining (for
the E hotspot) $C \sim 1.3 \times 10^{41}$ Hz$^4$, $p_i = 7.1$ per
cent: we would then expect the degree of polarization at 8 GHz to be
essentially zero (about $10^{-5}$ per cent) whereas in fact there is a
detection of polarization in this region at the $5\sigma$ level (Table
\ref{fltab}). The intrinsic polarization $p_i$ obtained here is also somewhat
low, which suggests that the model is not accurate even for the two
higher frequencies. However, the value of $C$, which is the critical
quantity for estimating physical parameters, simply depends
logarithmically on the ratio of the polarizations at the two
frequencies, and so is unlikely to be grossly wrong. Applying the
model, we can see from the {\it HST} data, assuming a cylindrical
symmetry, that $R \sim 5$ kpc: this would imply $(n_e B_\parallel)_f d
\approx 2 \times 10^{5}$ m$^{-3}$ nT kpc$^{-1}$. If we further assume
a simple model in which the differences are due to varying
orientations of the magnetic field in an approximately uniform-density
(though possibly low-filling-factor) medium, then, dropping numerical
factors of order unity, $n_e B d \approx 10^{5}$ m$^{-3}$ nT
kpc$^{-1}$. (If there are in fact variations in $n_e$, then the $n_e$
value derived from this equation is a weighted average.)

Unfortunately, none of these three parameters, $n_e$, $B$ and $d$, is
well constrained by existing observations. $B$ values of the order of
1 nT are energetically plausible from equipartition arguments, but
this estimate is good to an order of magnitude at best. We know that
$d < 1$ kpc for the model above to be either necessary or applicable;
the fact that the degree of polarization of the hotspots continues to
increase at higher radio resolutions (Table \ref{fltab}) suggests,
however, that at least some structure is present on scales $\sim 100$
pc.

The density is the most problematic quantity. Clark \etal\ (1998)
suggested that the density in the emission-line clouds in the nuclear
regions is $\sim 300$ cm$^{-3}$, but were unable to make measurements
near the hotspots. Electron densities $\sim 100$ cm$^{-3}$ are often
quoted for EELR. However, it is important to emphasise that it is
probably {\it not} the line-emitting material itself that is
depolarizing the radio source, if it has a density close to this
value. Baum \& Heckman (1989) estimate constraints on the density and
mass of the extended line-emitting material in 3C\,171, based on the
observed H$\alpha$ luminosity and assuming case B recombination: for a
filling factor of unity, they estimate $n_e = 0.2$ cm$^{-3}$. To
obtain a density $\sim 100$ cm$^{-3}$, a filling factor of $\sim
10^{-6}$ would be required: in other words, if the line-emitting
material has this density, it will have a very low covering fraction,
and cannot be responsible for depolarizing the radio source. For
filling factors close to unity, a single-phase medium with a density
close to the value given by Baum \& Heckman can still just about
explain the observations, requiring $d \sim 1$ kpc for $B = 1$ nT.
However, if the EELR consists of dense ($\sim 100$ cm$^{-3}$), warm
line-emitting clouds with a low filling factor, they cannot be
responsible for the observed depolarization. It must instead be due to
a second phase of the ISM. A plausible candidate is a second, hotter,
more diffuse ionized phase, which would be expected to be present in
shock-ionization models for the EELR: if the two phases are in
approximate pressure equilibrium, then the density of the hot phase
can be much lower than in the EELR clouds. If we take $d \approx 100$
pc and $B \approx 1$ nT, then the depolarization implies $n_e \approx
1$ cm$^{-3}$, which is a plausible density for the hot phase in the
above scenario. For pressure balance with the colder, denser EELR
clouds, that density would imply $T_{\rm hot} \sim 10^6$ K: of course,
we have no particular {\it a priori} reason to expect exact pressure
balance in this dynamic situation.

\begin{figure}
\epsfxsize \linewidth
\epsfbox{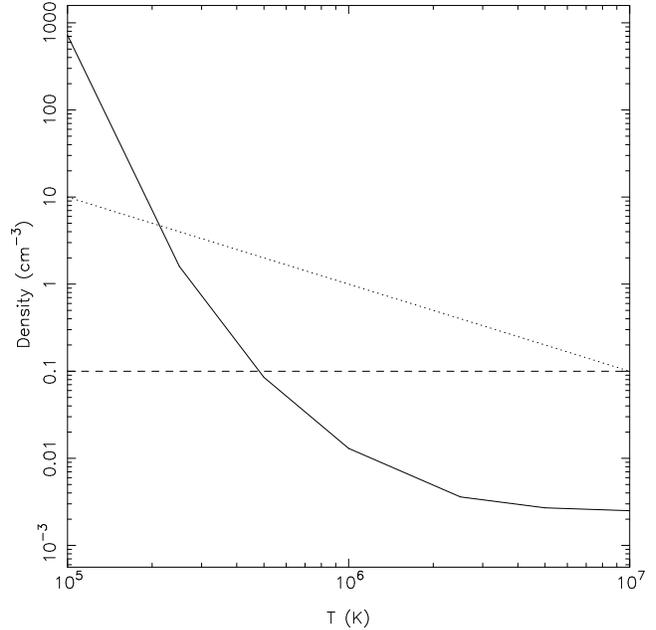}
\caption{Constraints on the properties of the depolarizing medium. The
solid line shows the upper limit on the electron density of the medium
as a function of temperature given the {\it ROSAT} upper limit
discussed in the text. Only regions of parameter space below and to
the left of the solid line are permitted by the X-ray upper limit:
higher densities or temperatures would produce more X-ray emission
than observed. The calculation was carried out using the {\sc mekal}
model within {\sc xspec}, assuming 0.5 solar abundances. The dashed
line shows the constraint imposed by the fact that the depolarizing
medium is not resolved ($d < 1$ kpc), assuming $B = 1$ nT. Only
regions of parameter space above this line are permitted by the data,
given equation \ref{depeq}. The exact position of the dashed line
depends on the magnitude of $B$, and so is uncertain, as discussed in
the text. The dotted line shows, for reference, the line of pressure
balance with a gas of $T = 10^4$ K, $n_e = 100$ cm$^{-3}$ (plausible
parameters for the EELR clouds).}
\label{nallow}
\end{figure}

The non-detection of 3C\,171 in a deep {\it ROSAT} X-ray observation
(Hardcastle \& Worrall 1999) sets quite strict limits on the
properties of any hot gas in the system. I derive a $3\sigma$ upper
limit on the absorbed {\it ROSAT}-band (0.1--2.4-keV) flux of the
region of interest of $\sim 3 \times 10^{-14}$ ergs cm$^{-2}$
s$^{-1}$. If we again assume cylindrical symmetry, with $R = 5$ kpc
 and $l \approx 50$ kpc, and consider a uniform medium, then Fig.\
\ref{nallow} shows the constraints on gas density and temperature
imposed by the X-ray limit together with the detection of
depolarization. It can be seen that plasma of the temperature
typically detected in the atmospheres of elliptical galaxies ($T \ga
10^7$ K) cannot be responsible for the depolarization: to avoid
detection in the X-ray, its density would have to be low ($10^{-2}$ --
$10^{-3}$ cm$^{-3}$), implying scales $d > 1$ kpc, which would be
resolved by the radio data. This conclusion is valid so long as $B$ is
not very much larger than 1 nT. The allowed region of parameter space
in Fig.\ \ref{nallow} is more or less consistent, given the large
uncertainties in all quantities involved (the calculation is
particularly uncertain at low temperatures), with the order of
magnitude estimate of the temperature of the `hot' depolarizing medium
given above: the X-ray data favour somewhat lower temperatures if $n_e
\sim 1$ cm$^{-3}$, $T \la 3 \times 10^5$ K. In the alternative scenario,
involving a low-density EELR region, we expect $T_{\rm EELR} \approx
10^4$ K, which of course is entirely consistent with the X-ray upper
limit.

Gas at temperatures around $10^6$ K can also in principle be detected
in the optical via coronal emission lines (e.g. Graney \& Sarazin
1990). If we adopt $n_e \approx 1$ cm$^{-3}$, then we can estimate
upper limits on the temperature of the gas given the non-detection of
any coronal lines in the spectra of Clark \etal . Graney \& Sarazin's
results suggest that the most significant line for a plasma in
ionization equilibrium (not necessarily a good model for the material
we are considering!) with $T < 10^6$ K is the [Fe X] 637.4 nm line. We
derive an upper limit on the flux of this line from Clark \etal's
detection of [O I] at 636.3 nm, giving $F_{\rm [Fe\ X]} \la 4 \times
10^{-16}$ ergs cm$^{-2}$ s$^{-1}$. For $n_e \approx 1$ cm$^{-3}$, given our
adopted geometry and the size of the extraction region of
Clark \etal\, we require the emissivity coefficient
$\Lambda_{\rm [Fe\ X]} \la 3
\times 10^{-27}$ ergs cm$^3$ s$^{-1}$. From Figure 1 of Graney \&
Sarazin, this corresponds to $T \la 6 \times 10^5$ K; higher
temperatures in the $10^{5.5}$--$10^{6.5}$ K region would have given
rise to detectable [Fe X] emission in the optical spectra. Thus the
optical data, like the X-ray data, favour temperatures somewhat lower
than $10^6$ K for the hot phase. Again, this conclusion is sensitive
to the choices of $d$ and $B$, which determine the adopted value of $n_e$.
Values of $B$ larger than 1 nT allow lower $n_e$ values and a hotter
plasma. In addition, it is sensitive to the metallicity and ionization
mechanism of the hot phase, which are unknown; the key point here is
simply to show that the presence of this phase is not ruled out by
existing optical observations.

To summarize, we are forced to one of two conclusions: either the EELR
regions are very much less dense than the `typical' numbers quoted for
$10^4$-K emission-line gas, and are distributed not in clumps but in a
uniform screen across the source; or, more probably, a second, less
dense phase of the ISM, with $n_e \sim 1$ cm$^{-3}$ and a temperature
likely to be around a few times $10^5$ K, is responsible for the
depolarization. This second phase is plausibly shocked warm diffuse
gas in the galaxy. In either case, the mass of the depolarizing
medium, assuming it is cospatial with the EELR, would be $\sim
10^{10}$ M$_{\sun}$.

\section*{Acknowledgements}

The National Radio Astronomy Observatory is a facility of the National
Science Foundation operated under cooperative agreement by Associated
Universities, Inc. I am grateful to Michael Rupen for help with the
special observing modes of the VLA used in this project, and the VLA
analysts for help in implementing them in an {\sc observe} file. I thank
Malcolm Bremer for useful discussions about the nature of EELR, and
Clive Tadhunter for supplying me with the {\it HST} image shown in
section 1.

\bsp

\end{document}